# LogitMat : Zeroshot Learning Algorithm for Recommender Systems without Transfer Learning or Pretrained Models

Hao Wang
Ratidar Technologies LLC
Beijing, China
haow85@live.com

*Abstract*— Recommender system is adored in the internet industry as one of the most profitable technologies. Unlike other sectors such as fraud detection in the Fintech industry, recommender system is both deep and broad. In recent years, many researchers start to focus on the cold-start problem of recommender systems. In spite of the large volume of research literature, the majority of the research utilizes transfer learning / meta learning and pretrained model to solve the problem. Although the researchers claim the effectiveness of the approaches, everyone of them does rely on extra input data from other sources. In 2021 and 2022, several zeroshot learning algorithm for recommender system such as ZeroMat, DotMat, PoissonMat and PowerMat were invented. They are the first batch of the algorithms that rely on no transfer learning or pretrained models to tackle the problem. In this paper, we follow this line and invent a new zeroshot learning algorithm named LogitMat. We take advantage of the Zipf Law property of the user item rating values and logistic regression model to tackle the cold-start problem and generate competitive results with other competing techniques. We prove in experiments that our algorithm is fast, robust and effective.

I. INTRODUCTION

Recommender system is a deep and broad research field that has significant impact on our society. Everyone of us is influenced by recommender system at least once in our lifetime : TikTok users enjoy video recommendation; Pandora / Netease Music / QQ Music users enjoy music recommendation; Netflix / iQiYi / YouTube watchers enjoy video recommendation; ByteDance readers enjoy news recommendation; Amazon / JD / Pin Duoduo consumers enjoy product recommendation. Recommender system technologies have propelled our internet products to a new height at which no modern day person could escape the influence of the technology. Since the year of 2012, with the advent of deep learning, researchers have shifted the research focus of recommender system from shallow models to deep neural networks. People have spent billions of money on R&D of the technologies and the effect is significant in both academia and the industry.

The earliest recommender system technology is collaborative filtering, which was invented more than 20 years ago. After the emergence of Amazon and Netflix and other companies that emphasize the application of the technology, a series of new approaches have been invented such as matrix factorization, learning to rank, and deep neural networks. The major task of the field is to boost the technical accuracy of the algorithm. Millions of different versions of recommender systems have been tested since its debut, and many classic models remain in circulation up to today.

In spite of the long-term efforts of researchers and industrial practitioners, there remain intrinsic problems of recommender systems that continue to bug the minds of scientists and researchers. One such problem is the cold-start problem. The cold-start problem refers to the issue arises when a new user or a new item emerges in the recommender system input data pool. Every recommender system faces the cold-start problem during its entire lifecycle. Popular techniques to tackle this problem include heuristics, product redesign, transfer learning / meta learning with pretrained model, etc. Transfer learning and meta learning approaches are most technical approaches in the field. However, both techniques require side information or input from some data sources. In the year of 2021, a new zeroshot learning algorithm named ZeroMat [1] was proposed to solve the problem - the algorithm relies on Zipf distribution of the user item rating values and the probabilistic matrix factorization model. The algorithm outperforms the classic matrix factorization if the parameters are carefully chosen. In the following year, ZeroMat's author proposes DotMat [2], PoissonMat [3] and PowerMat [4], all of which are zeroshot learning algorithms that consume no user item rating data at all.

In this paper, we follow the line of thoughts initiated by ZeroMat and propose a new algorithm named LogitMat. The algorithm relies on the Zipf law property of the user item rating dataset and the redesign of the logistic regression model. We prove in the experiments the superiority of our algorithms and conclude that our algorithm is fast, robust and effective in modern day recommendation contexts.

## II. RELATED WORK

Recommender system contains a broad spectrum of technologies such as collaborative filtering (user-based collaborative filtering [5] / item-based collaborative filtering [6]), matrix factorization (probabilistic matrix factorization [7] / SVD++ [8]), learning to rank (Bayesin Personalized Ranking [9] / Collaborative Less is More Filtering [10]), deep neural networks (DCN [11] / DeepFM [12] / Wide & Deep [13]). We provide an overview of the recommender system technologies in this section.

User-based and item-based collaborative filtering are the most primitive forms of modern day recommender system approaches. In 2018, Wang et. al. [14] propose a set of combinatorial tools to analyze the fairness problem related to the 2 approaches. The research paper is the first of its kind that provides mathematically sound analysis of popularity bias problem fo the field. A more recent discovery of the field is Kernel-CF algorithm [15] invented in 2022, which builds connection among nonparametric statistics, social network analysis and recommender systems.

The most successful recommender system paradigm is matrix factorization. The most significant milestone in the field is the probabilistic matrix factorization [7]. The algorithm models the matrix factorization framework as an MAP problem that is extensible to many particular inventions. SVD++ [8] and timeSVD [16] are specific examples of a theoretical framework named SVDFeature [17] which builds feature engineering into the matrix factorization framework. In 2021, a new algorithm named MatRec [18] was invented as a special case of SVDFeature to tackle the fairness problem of recommender systems. In the same year, Zipf Matrix Factorization method was proposed to solve the popularity bias problem by adding a regularization term that penalizes the Degree of Matthew Effect.

Learning to Rank was invented around the year of 2010. The most classic examples of Learning to Rank algorithms are Bayesian Personalized Ranking [9] and Collaborative Less is More Filtering [10]. Learning to Rank algorithms could be classified as Point-wise, Pair-wise and List-wise approaches. In 2022, Pareto Pairwise Ranking [19] was proposed to attack both the accuracy and fairness problems related to the learning to rank problem. Other fairness-related learning to rank algorithms could be found at SIGIR and other top research venues [20] [21] [22].

Deep neural networks is the new buzz word today. Facebook introduced its own deep neural network-based recommender system DLRM [23] a couple of years ago. DeepFM [12] and Wide & Deep [13] are both popularly adopted techniques in the Chinese internet companies. Deep Matrix Factorization [24] is the deep learning era matrix factorization algorithm.

Applications of recommender systems in the industry include Netease's Online Dating Services [25], Baidu's Question & Answering Product [26] [27], etc. Recommender system has also intrigued researchers to work in other fields, such as computational cultural studies, with the pioneering publication published in 2022 [28], discussing how zeroshot learning algorithm could explain the lock-in state effect of human culture / civilization evolution.

Other than accuracy enhancement, there are research literature on other aspects of recommender system technologies. One such important research subfield is fairness. In 2017, Google researcher Ed. Chi published a paper on focused learning [29], talking about how to penalize matrix factorization paradigm to enhance fairness performance. In 2022, KL-Mat [30] was invented to solve the popularity bias problem using KL-Divergence penalty. A set of fairness metrics [31] for recommender systems were proposed in 2022 using extreme value theory in statistics.

Another hot research topic in the field of recommender systems is zeroshot learning. Zeroshot learning approaches are used to solve the cold-start problem, with application of transfer learning and meta learning [32][33][34]. A significant breakthrough took place in the year of 2021 and 2022 by introduction of ZeroMat [1], DotMat [2], PoissonMat[3] and PowerMat [4]. The 4 algorithms rely on no user item rating value data for recommendation and personalizatoin. Explainable recommender system is also a rising research topic in recent years. One notable publication is Wang [35], which explains and designs recommender system using geometry.

## III. LOGITMAT FORMULATION

We build our LogitMat algorithm upon the logistic regression model. Formally, we redesign the logistic regression model to solve the recommendation problem. As well known in the machine learning community, logistic regression is defined in the following way :

$$P(C|X) = \begin{cases} \dfrac{\exp(w \cdot x)}{1 + \exp(w \cdot x)}, & C = 0 \\ \dfrac{1}{1 + \exp(w \cdot x)}, & C = 1 \end{cases}$$

Another algorithmic framework that we need in our problem formulation is matrix factorization. Formally, matrix factorization is defined as below :

$$L = \sum_{i=1}^{n} \sum_{j=1}^{m} (R_{i,j} - U_i^T \cdot V_j)^2$$

Matrix factorization paradigm could be viewed as a dimensionality reduction problem that consumes only linear space complexity for quadratic space complexity data. The formal theoretical foundation for matrix factorization is probabilistic matrix factorization (PMF) [7], which models the matrix factorization problem as an MAP problem.

As expained in ZeroMat [1], DotMat [2] and other publications [3][21], the distribution of user item rating values for recommender systems follow Zipf Law. In other words, for a move review system, if the user rating scale is between 1 and 5, with 5 being the highest score, then the movie with n star rating value in collective opinion has audience number proportional to n. To elaborate, let's say *Movie A* has a rating 5 on our movie review website, and *Movie B* has a rating 4 on our movie review website, then the ratio between *Movie A* watcher number and *Movie B* watcher number is 5: 4. We assume our recommender system dataset has a user value rating scale 1 to m, with m being the highest score, and define the algorithm in the following way :

$$\frac{R_{i,j}}{R_{max}} = \begin{cases} \dfrac{exp(W_i^T \cdot Z_j \cdot U_i^T \cdot V_j)}{1 + exp(W_i^T \cdot Z_j \cdot U_i^T \cdot V_j)} \text{ , with probability } \dfrac{\sum_{i=1}^{K-1} n_i}{\sum_{i=1}^{K} n_i} \\ \dfrac{1}{1 + exp(W_i^T \cdot Z_j \cdot U_i^T \cdot V_j)} \text{ , with probability } \dfrac{1}{\sum_{i=1}^{K} n_i} \end{cases}$$

To transform the algorithm into a solvable loss function, we have :

$$L = \begin{cases} \left(\dfrac{U_i^T \cdot V_j}{R_{max}} - \dfrac{exp(W_i^T \cdot Z_j \cdot U_i^T \cdot V_j)}{1 + exp(W_i^T \cdot Z_j \cdot U_i^T \cdot V_j)}\right)^2 \text{ , with probability } \dfrac{\sum_{i=1}^{K-1} n_i}{\sum_{i=1}^{K} n_i} \\ \left(\dfrac{U_i^T \cdot V_j}{R_{max}} - \dfrac{1}{1 + exp(W_i^T \cdot Z_j \cdot U_i^T \cdot V_j)}\right)^2 \text{ , with probability } \dfrac{1}{\sum_{i=1}^{K} n_i} \end{cases}$$

In this formulation, W and Z are coefficient vectors, U is the user feature vector, and V is the item feature vector. $n_i$ is the i-th largest user item rating value in the recommender system.

The reason why we set probability for logistic expressions is because we eliminate features from the logistic regression model so our algorithm will become input data free, and therefore is a zeroshot learning algorithm. In the classic logistic regression model, we have 2 different forms of logit expressions, so their sum would be 1. In our case, we do not have training data, so we can not set logit expression forms for each label class analytically and deterministically, so we choose a probabilistic method to select which logit expression to use. The reason why we choose differnt forms of logit functions with different probabilities is because we want the sum of the different logit expressions to be 1.

We apply Stochastic Gradient Descent (SGD) algorithm to solve for the optimal values of the parameters, and obtain the following formulas :

$$\begin{cases} \text{With probability } \dfrac{\sum_{i=1}^{K-1} n_i}{\sum_{i=1}^{K} n_i} : \begin{cases} \dfrac{\partial L}{\partial U_i} = \dfrac{1}{R_{max}} V_j - \left(\dfrac{t_0 t_1}{t_2} V_j - \dfrac{t_0 t_1^2}{t_2^2} V_j\right), t_0 = W_i^T \cdot Z_j, t_1 = exp(t_0 U_i^T \cdot V_j), t_2 = 1 + t_1 \\ \dfrac{\partial L}{\partial V_j} = \dfrac{1}{R_{max}} U_i - \left(\dfrac{t_0 t_1}{t_2} U_i - \dfrac{t_0 t_1^2}{t_2^2} U_i\right), t_0 = W_i^T \cdot Z_j, t_1 = exp(t_0 U_i^T \cdot V_j), t_2 = 1 + t_1 \\ \dfrac{\partial L}{\partial W_i} = -\left(\dfrac{t_0 t_1}{t_2} Z_j - \dfrac{t_0 t_1^2}{t_2^2} Z_j\right), t_0 = U_i^T \cdot V_j, t_1 = exp(t_0 W_i^T \cdot Z_j) \\ \dfrac{\partial L}{\partial Z_j} = -\left(\dfrac{t_0 t_1}{t_2} W_i - \dfrac{t_0 t_1^2}{t_2^2} W_i\right), t_0 = U_i^T \cdot V_j, t_1 = exp(t_0 W_i^T \cdot Z_j) \end{cases} \\ \text{With probability } \dfrac{1}{\sum_{i=1}^{K} n_i} : \begin{cases} \dfrac{\partial L}{\partial U_i} = \dfrac{2 t_4}{R_{max}} V_j + \dfrac{2 t_1 t_2 t_4}{t_3^2} V_j, t_0 = U_i^T \cdot V_j, t_1 = W_i^T \cdot Z_j, t_2 = exp(t_0 t_1), t_3 = 1 + t_2, t_4 = \dfrac{t_0}{R_{max}} - \dfrac{1}{t_3} \\ \dfrac{\partial L}{\partial V_j} = \dfrac{2 t_4}{R_{max}} U_i + \dfrac{2 t_1 t_2 t_4}{t_3^2} U_i, t_0 = U_i^T \cdot V_j, t_1 = W_i^T \cdot Z_j, t_2 = exp(t_0 t_1), t_3 = 1 + t_2, t_4 = \dfrac{t_0}{R_{max}} - \dfrac{1}{t_3} \\ \dfrac{\partial L}{\partial W_i} = \dfrac{2 t_0 t_1 \left(\dfrac{t_0}{R_{max}} - \dfrac{1}{t_2}\right)}{t_2^2} Z_j, t_0 = U_i^T \cdot V_j, t_1 = exp(t_0 W_i^T \cdot Z_j), t_2 = 1 + t_1 \\ \dfrac{\partial L}{\partial Z_j} = \dfrac{2 t_0 t_1 \left(\dfrac{t_0}{R_{max}} - \dfrac{1}{t_2}\right)}{t_2^2} W_i, t_0 = U_i^T \cdot V_j, t_1 = exp(t_0 W_i^T \cdot Z_j), t_2 = 1 + t_1 \end{cases} \end{cases}$$

## IV. EXPERIMENTS

We test our algorithm on MovieLens 1 Million Dataset [36] and LDOS-CoMoDa Dataset [37]. MovieLens 1 Million Dataset comprises of 6040 users and 3952 items, while LDOS-CoMoDa Dataset is a smaller dataset but contains contextual information for movie recommendation. To evaluate the performance of LogitMat, we compare the algorithm with 7 other modern-day algorithms, namely ZeroMat [1], DotMat [2], PoissonMat [3], ParaMat [39], Pareto Pairwise Ranking [19], Skellam Rank [38] and classic matrix factorization. The evaluation metric we use for comparison are Mean Absolute Error (MAE) and Degree of Matthew Effect [40].

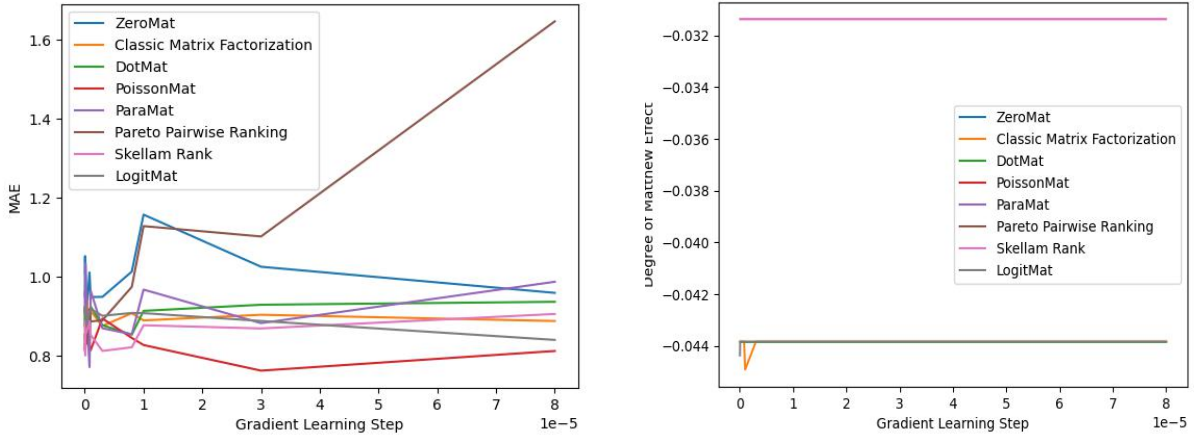

Fig. 1 LogitMat comparison with other algorithms on MovieLens Dataset

From Fig.1 we notice that LogitMat ranks No.1 in MAE comparison and ranks No.2 in Degree of Matthew Effect comparison. This proves the effectiveness of the algorithm. Notice that Classic Matrix Factorization and Skellam Rank are full data algorithms. Zeroshot learning algorithm LogitMat outperforms the full data algorithms and other zeroshot learning algorithms, including the best performing one (PoissonMat) so far.

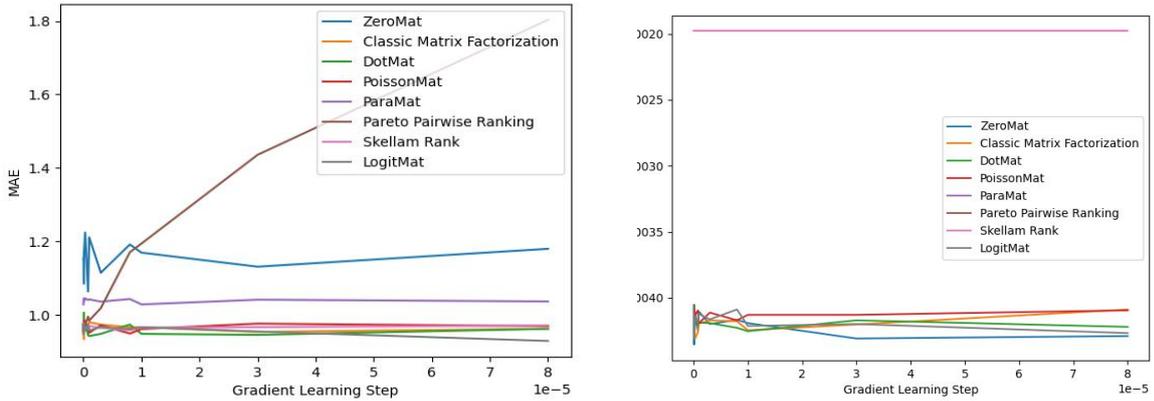

Fig. 2 LogitMat comparison with other algorithms on LDOS-CoMoDa Dataset

From Fig. 2 we notice that LogitMat ranks No.3 in MAE comparison when the gradient learning step becomes larger, and No.4 in the overall MAE experiment. However, when it comes to both accuracy and robustness, LogitMat ranks No.3 in all algorithms, and No.2 in Zeroshot Learning algorithm comparison. In fairness metric comparison experiments, LogitMat ranks No.2 in all experiments, and No.1 in Zeroshot Learning algorithm comparison.

## V. Conclusion

Recommender system experiences rapid progress as a technical field in both the academia and industry. However, intrinsic problems such as cold-start issues have bugged the researchers and developers for decades. In 2021 and 2022, the author of this paper proposed 4 zeroshot learning algorithms that solve the problem without input data. In this paper, the same author proposed a new algorithm named LogitMat that also solves this problem without input data. The performance of the algorithm is competitive with the best performing zeroshot learning algorithms, as well as full data algorithms such as matrix factorization and Skellam Rank [38].

In future work, we would like to explore other models that can solve the cold-start problem for recommender systems. We would also like to extend our experiences into the field of natural language processing and computational social sciences (which we already published 1 paper [28] and initiated a new field we'd like to call *Computational Cultural Studies*).

## VI. References

[1] H.Wang, "ZeroMat: Solving Cold-start Problem of Recommender System with No Input Data", IEEE 4th International Conference on Information Systems and Computer Aided Education (ICISCAE), 2021

[2] H.Wang, "DotMat: Solving Cold-start Problem and Alleviating Sparsity Problem for Recommender Systems", ICET, 2022

[3] H. Wang, "PoissonMat: Remodeling Matrix Factorization using Poisson Distribution and Solving the Cold Start Problem without Input Data", MLISE, 2022

[4] H. Wang, "PowerMat: context-aware recommender system without user item rating values that solves the cold-start problem", AIID, 2022

[5] Z.D. Zhao, M.S. Shang, "User-Based Collaborative-Filtering Recommendation Algorithms on Hadoop. Third International Conference on Knowledge Discovery and Data Mining", 2010

[6] B.Sarwar, G.Karypis, et.al. "Item-based collaborative filtering recommendation algorithms", WWW, 2001

[7] R.Salakhutdinov, A.Mnih, "Probabilistic Matrix Factorization", Proceedings of the 20th International Conference on Neural Information Processing Systems, 2007

[8] S. Wang, G. Sun, Y. Li, "SVD++ Recommendation Algorithm Based on Backtracking", Information, 2020

[9] S. Rendle, C. Freudenthaler, et. al. "BPR: Bayesian Personalized Ranking from Implicit Feedback", Proceedings of the Twenty-Fifth Conference on Uncertainty in Artificial Intelligence, 2009

[10] Y. Shi, A. Karatzoglou, et. al. , "CLiMF: learning to maximize reciprocal rank with collaborative less-is-more filtering", Proceedings of the sixth ACM conference on Recommender systems, 2012

[11] R. Wang, R. Shivanna, et. al., "DCN V2: Improved Deep & Cross Network and Practical Lessons for Web-scale Learning to Rank Systems", WWW, 2021

[12] H. Guo, R. Tang, et. al., "DeepFM: A Factorization-Machine based Neural Network for CTR Prediction", IJCAI, 2017

[13] H. Cheng, L.Koc, et.al. , "Wide & Deep Learning for Recommender Systems", DLRS, 2016

[14] H. Wang, "Quantitative Analysis of Matthew Effect and Sparsity Problem in Recommender Systems", ICCCBDA, 2018

[15] H. Wang, "Kernel-CF: Collaborative filtering done right with social network analysis and kernel smoothing", CAIBDA, 2022, to appear

[16] Y. Koren, "Collaborative filtering with temporal dynamics", Communications of the ACM 53.4 (2010): 89-97, 2010

[17] T.Chen, W.Zhang, Q.Lu, et.al., "SVDFeature: A Toolkit for Feature-based Collaborative Filtering", Journal of Machine Learning Research, 2012

[18] H. Wang, "MatRec: Matrix Factorization for Highly Skewed Dataset. 3rd International Conference on Big Data Technologies", 2020

[19] H. Wang, "Pareto pairwise ranking for fairness enhancement of recommender systems", CISAT, 2022

[20] M.Zehlike, C.Castillo. "Reducing Disparate Exposure in Ranking: A Learning to Rank Approach.", SIGIR, 2020

[21] H.Yadav, Z.Du, T.Joachims. "Fair Learning-to-Rank from Implicit Feedback.", SIGIR, 2020

[22] M.Morik, A.Singh, J.Hong, T.Joachims. "Controlling Fairness and Bias in Dynamic Learning-to-Rank", SIGIR, 2020

[23] M.Naumov, D.Mudigere, et. al. , "Deep Learning Recommendation Model for Personalization and Recommendation Systems", CoRR, 2019

[24] H. Xue, X. Dai, J. Zhang, et. al. "Deep Matrix Factorization Models for Recommender Systems", Proceedings of the Twenty-Sixth International Joint Conference on Artificial Intelligence, 2017

[25] C. Dai, F. Qian, et. al. "A personalized recommendation system for NetEase dating site", VLDB Endowment, 2014

[26] Q. Liu, T. Chen, J. Cai, et. al. , " Enlister: baidu's recommender system for the biggest chinese Q&A website", ACM RecSys, 2012

[27] T. Chen, J. Cai, H. Wang, et. al. , "Instant expert hunting: building an answerer recommender system for a large scale Q&A website", ACM SAC, 2014

[28] H. Wang, "Is Human Culture Locked by Evolution?", MHEHD, 2022

[29] A. Beutel, E. Chi, C. Cheng. et. al. , "Beyond Globally Optimal: Focused Learning for Improved Recommendations", WWW, 2017

[30] H. Wang, "KL-Mat: Fair Recommender System via Information Geometry", International Conference on Wireless Communication and Sensor Networks (icWCSN), 2022

[31] H. Wang, "Fairness Metrics for Recommender Systems", icWCSN, 2022

[32] Y. Du, S. Rendle, et. al. , "Zero-Shot Heterogeneous Transfer Learning from Recommender Systems to Cold-Start Search Retrieval", CIKM, 2020

[33] Y. Zheng, S. Liu, et.al., "Cold-start Sequential Recommendation via Meta Learner", AAAI, 2021

[34] M. Vartak, Arvind Thiagarajan, Conrado Miranda, "A Meta-Learning Perspective on Cold-Start Recommendations for Items", NIPS, 2017

[35] H. Wang, "Fair Recommendation by Geometric Interpretation and Analysis of Matrix Factorization", RAIIE, 2022

[36] MovieLens Dataset. https://grouplens.org/datasets/movielens/

[37] ODIĆ, Ante, TKALČIČ, Marko, TASIČ, Jurij F., KOŠIR, Andrej, 2013, Predicting and Detecting the Relevant Contextual Information in a Movie-Recommender System, Interacting with Computers, Volume 25, Issue 1

[38] H. Wang, "Skellam Rank: Fair Learning to Rank Algorithm based on Poisson Process and Skellam Distribution for Recommender Systems", AIBT, 2023, to appear

[39] H. Wang, "Fair Recommendation by Geometric Interpretation and Analysis of Matrix Factorization", RAIIE, 2022

[40] H. Wang, "Zipf Matrix Factorization : Matrix Factorization with Matthew Effect Reduction", ICAIBD, 2021